\documentclass{llncs}

\usepackage{hyperref}
\usepackage{xspace}
\usepackage{amssymb}
\usepackage{amsmath}

\usepackage{keyjavadl}
\usepackage{keybook}

\newcommand{\openLoopScope}{\ensuremath{\circlearrowright}}
\newcommand{\closeLoopScope}{\ensuremath{\circlearrowleft}}

\begin{document}

\title{Technical Report: \\ Using Loop Scopes with \java{for}-%
Loops}
\author{Nathan Wasser \and Dominic Steinh\"ofel}
\institute{Technische Universit\"at Darmstadt, Department of Computer Science,\\
  64289 Darmstadt, Germany\\
\email{$\{$wasser,steinhoefel$\}$@cs.tu-darmstadt.de}\\
\url{https://www.informatik.tu-darmstadt.de/se}}
\date{\today}
\maketitle

\begin{abstract}
Loop scopes have been shown to be a helpful tool in creating sound loop invariant rules which do not require program transformation of the loop body.
Here we extend this idea from while-loops to for-loops and also present sound loop unrolling rules for while- and for-loops, which require neither program transformation of the loop body, nor the use of nested modalities.
This approach allows for-loops to be treated as first-class citizens -- rather than the usual approach of transforming for-loops into while-loops -- which makes semi-automated proofs easier to follow for the user, who may need to provide help in order to finish the proof.

\keywords{Theorem proving \and Loop invariants}
\end{abstract}

\section{Introduction}

\emph{Loop scopes} were introduced in~\cite{Wasser16}. These allow a sound loop invariant rule (which does not require program transformation of the loop body) for programs with non-standard control-flow. In~\cite{SteinhofelW17} it was shown that an implementation of this new loop invariant rule in \KeY~\cite{KeYBook2016} also decreases proof size when compared to the existing rule.

The loop invariant rule using loop scopes was proposed only for \java{while} loops. Here we briefly sketch how loop invariant rules for
\java{for} loops using loop scopes can be introduced to avoid program transformation of the loop bodies.

\section{Implementation of Loop Scope Rules in \KeY}

When the rules for loop scopes were imlemented in \KeY some changes were made.
Among others the implemented loop invariant rule for while loops using loop scopes is as follows:

\[
\seqRuleW{loopInvariantWhile}%
{\sequent{}{\upl \mathcal{U} \upr \mathit{Inv}} \\
\sequentb{\Gamma, \upl \mathcal{U}' \upr \mathit{Inv}}{\upl \mathcal{U}'~||~\java{x}\upd\text{TRUE} \upr {\quad\ } \\
\qquad\qquad\qquad\quad \dlboxf{$\pi$ $\openLoopScope_{\java{x}}$ if ($nse$) $l_1:\ldots l_n:$ \{ p x = false; \} $\closeLoopScope$ $\omega$}{\quad\ }} \\
\hfill ((\java{x} \doteq \text{TRUE} \rightarrow \phi)\ \&\ (\java{x} \doteq \text{FALSE} \rightarrow \mathit{Inv})), \Delta
}%
{\sequent{}{\upl \mathcal{U} \upr \dlboxf{$\pi$ $l_1:\ldots l_n:$ while ($nse$) p $\omega$}{\phi}}}
\]

In the context of the loop invariant rule these changes to the loop scope rules do not impact the semantics. However, now continuing a loop normally only implicitly sets the loop scope index variable to false, rather than explicitly calling \java{continue}.

Another change is that the definition of the rule for the empty loop scope in \KeY no longer matches the original rule of~\cite{SteinhofelW17}, where the index was set to true. Now, the index is initialized to true by the loop invariant rule, and the rule processing the empty loop scope checks for its value to decide how to continue:

\[
\seqRuleW{emptyIndexedLoopScope}%
{\sequent{}{\upl \mathcal{U} \upr $(\IfThenElse{\java{x}\doteq\TRUE}{\dlboxf{$\pi$ $\omega$}{\varphi}}{\dlbox{}{\varphi}})$}}%
{\sequent{}{\upl \mathcal{U} \upr \dlboxf{$\pi$ $\openLoopScope_{\java{x}}$ $\closeLoopScope$ $\omega$}{\varphi}}}
\]

In order to keep the semantics equivalent for normal loop continuation and explicit loop continuation via \java{continue} statement inside a loop scope, we consider the following formulas\footnote{The implemented loop invariant rule for while loops cannot result in these formulas.}:

\begin{align}
  \upl \java{b} \doteq \text{TRUE} \upr \dlboxf{$\openLoopScope_{\java{x}}$ if (b) \{ b = false; x = false; \} if (!x) \{ p \} $\closeLoopScope$}{\phi} \label{eq:x_false_1} \\
  \upl \java{b} \doteq \text{TRUE} \upr \dlboxf{$\openLoopScope_{\java{x}}$ if (b) \{ b = false; continue; \} if (!x) \{ p \} $\closeLoopScope$}{\phi} \label{eq:continue_1}
\end{align}

Simplifying these formulas leads to:

\begin{align}
  \upl \java{b} \doteq \text{FALSE}~||~\java{x} \doteq \text{FALSE} \upr \dlboxf{if (!x) \{ p \}}{\phi} \label{eq:x_false_2} \\
  \upl \java{b} \doteq \text{FALSE} \upr \dlboxf{$\openLoopScope_{\java{x}}$ continue; if (!x) \{ p \} $\closeLoopScope$}{\phi} \label{eq:continue_2}
\end{align}

Simplifying \eqref{eq:x_false_2} further leads to the formula
$\upl \java{b} \doteq \text{FALSE}~||~\java{x} \doteq \text{FALSE} \upr \dlboxf{p}{\phi}$.
As we want the semantics to be equivalent, simplifying \eqref{eq:continue_2} needs to result in the same formula. The easiest solution is to ensure that simplifying \eqref{eq:continue_2} leads to \eqref{eq:x_false_2}, by replacing the original rule in~\cite{Wasser16} with the following:

\[
\seqRule{continueIndexedLoopScope}%
{\sequent{}{\upl \mathcal{U} \upr \dlboxf{$\pi$ $\openLoopScope_{\java{x}}$ x = false; p $\closeLoopScope$ $\omega$}{\varphi}}}%
{\sequent{}{\upl \mathcal{U} \upr \dlboxf{$\pi$ $\openLoopScope_{\java{x}}$ continue; p $\closeLoopScope$ $\omega$}{\varphi}}}
\]

The difference is that in the original rule the program fragment \java{p} is thrown away, while here it is kept (the context is also not deleted, which works well because of the described change to \textsf{emptyIndexedLoopScope}). This has no influence whatsoever on the existing loop invariant rule, which would only ever lead to applications of the $\mathsf{continueIndexedLoopScope}$ rule with an empty \java{p} (since the body is contained in a block, and the remainder of a block is discarded by the $\mathsf{blockContinue}$ rule). However, it allows us to treat for-loops. In fact, the example formulas \eqref{eq:x_false_1} and \eqref{eq:continue_1} use loop scopes to model the following two programs%
\footnote{As will be shown later, this is not quite accurate, as \java{p} is an expression list inside the programs, but a program fragment in the formulas. While an expression list is not a program fragment, it can easily be converted into one.}:

\begin{align}
 &\java{b = true; for (;b;p) \{ b = false; \}} \\
 &\java{b = true; for (;b;p) \{ b = false; continue; \}}
\end{align}

\section{Loop Invariant Rules using Loop Scopes}

In order to prove that the loop invariant of a \java{for} loop initially holds, we must first reach the ``initial'' entry point of the loop. This is the point after full execution of the loop initializer. We therefore introduce the following rule to pull out the loop initializer of a \java{for} loop, where $init'$ is a statement list equivalent to the loop initializer $init$:

\[
\seqRuleW{pullOutLoopInitializer}%
{\sequent{}{\upl \mathcal{U} \upr \dlboxf{$\pi$ \{ $init'$ $l_1:\ldots l_n:$ for (; $guard$; $upd$) p \} $\omega$}{\phi}}
}%
{\sequent{}{\upl \mathcal{U} \upr \dlboxf{$\pi$ $l_1:\ldots l_n:$ for ($init$; $guard$; $upd$) p $\omega$}{\phi}}}
\]

The following loop invariant rule can then be applied to \java{for} loops without loop initializers, where $upd'$ is a statement list equivalent to the expression list $upd$, and $guard'$ is an expression equivalent to $guard$ (\java{true}, if $guard$ is empty):

\begin{align*}
\seqRuleW{loopInvariantFor}%
{\sequent{}{\upl \mathcal{U} \upr \mathit{Inv}} \\
\sequentb{\Gamma, \upl \mathcal{U}' \upr \mathit{Inv}}{\upl \mathcal{U}'~||~\java{x}\upd\text{TRUE} \upr \\
\qquad\qquad\qquad\quad [\java{$\pi$ $\openLoopScope_{\java{x}}$ if ($guard'$) $l_1:\ldots l_n:$ \{ p x = false; \}} {\qquad\ \ } \\
\hfill \java{if (!x) \{ x = true; $upd'$ x = false; \} $\closeLoopScope$ $\omega$}]} \\
\hfill ((\java{x} \doteq \text{TRUE} \rightarrow \phi)\ \&\ (\java{x} \doteq \text{FALSE} \rightarrow \mathit{Inv})), \Delta
}%
{\sequent{}{\upl \mathcal{U} \upr \dlboxf{$\pi$ $l_1:\ldots l_n:$ for (; $guard$; $upd$) p $\omega$}{\phi}}}
\end{align*}

Execution of $upd'$ must be wrapped (\java{\{ x = true; $upd'$ x = false; \}}) to ensure that an exception thrown in $upd'$ causes \java{x} to be set to \java{true} as required. 


\section{Loop Unwinding Rules using Loop Scopes}

Using this new idea we can also create a loop unwinding rule for \java{while} loops which does not require nested modalities as the rule in~\cite{Wasser16} does:

\[
\seqRuleW{unwindWhileLoop}%
{\sequent{}{\upl \mathcal{U}~||~\java{x}\upd\text{TRUE}~||~\java{cont}\upd\text{FALSE} \upr {\quad\ } \\
\qquad\quad[\java{$\pi$ $\openLoopScope_{\java{x}}$ if ($nse$) $l_1:\ldots l_n:$ \{ p x = false; \}} \\
\qquad\qquad\qquad\,\java{if (!x) \{ x = true; cont = true; \} $\closeLoopScope$} \\
\qquad\qquad\ \java{if (cont) $l_1:\ldots l_n:$ while ($nse$) p $\omega$}]{\phi}}
}%
{\sequent{}{\upl \mathcal{U} \upr \dlboxf{$\pi$ $l_1:\ldots l_n:$ while ($nse$) p $\omega$}{\phi}}}
\]

The idea here is to execute one loop iteration within a loop scope with a continuation that tracks that the loop should be continued and then exits the loop scope. Based on the value of the tracking variable we then either re-enter the loop or continue with the rest of the program.


We can also introduce a loop unwinding rule for the \java{for} loop:

\[
\seqRuleW{unwindForLoop}%
{\sequent{}{\upl \mathcal{U}~||~\java{x}\upd\text{TRUE}~||~\java{cont}\upd\text{FALSE} \upr {\quad\ } \\
\qquad\quad[\java{$\pi$ $\openLoopScope_{\java{x}}$ if ($guard'$) $l_1:\ldots l_n:$ \{ p x = false; \}} \\
\qquad\qquad\qquad\,\java{if (!x) \{ x = true; $upd'$ cont = true; \} $\closeLoopScope$} \\
\qquad\qquad\ \java{if (cont) $l_1:\ldots l_n:$ for (; $guard$; $upd$) p $\omega$}]{\phi}}
}%
{\sequent{}{\upl \mathcal{U} \upr \dlboxf{$\pi$ $l_1:\ldots l_n:$ for (; $guard$; $upd$) p $\omega$}{\phi}}}
\]

It is important that $upd'$ is executed after setting \java{x} to \java{true} and before setting \java{cont} to \java{true}, to ensure that an exception thrown in $upd'$ leaves \java{x} set to \java{true} while \java{cont} remains \java{false}. 

\bibliographystyle{splncs04}
\bibliography{main}

\end{document}